\documentclass[sigconf, nonacm]{acmart}
\AtBeginDocument{%
  \providecommand\BibTeX{{%
    \normalfont B\kern-0.5em{\scshape i\kern-0.25em b}\kern-0.8em\TeX}}}

\setcopyright{acmcopyright}
\copyrightyear{2021}
\acmYear{2021}
\acmDOI{XXXXXXX.XXXXXXX}

\begin{document}

\title{Online Aggregation based Approximate Query Processing: A Literature Survey}

\author{Pritom Saha Akash}
\email{pakash2@illinois.edu}
\affiliation{%
  \institution{University of Illinois at Urbana-Champaign}
  \city{Champaign}
  \country{USA}
}
\author{Wei-Cheng Lai}
\email{wclai2@illinois.edu}
\affiliation{%
  \institution{University of Illinois at Urbana-Champaign}
  \city{Champaign}
  \country{USA}
}

\author{Po-Wen Lin}
\email{powenwl2@illinois.edu}
\affiliation{%
  \institution{University of Illinois at Urbana-Champaign}
  \city{Champaign}
  \country{USA}
}


\begin{abstract}
  In the current world, OLAP (Online Analytical Processing) is used intensively by modern organizations to perform ad hoc analysis of data, providing insight for better decision making. Thus, the performance for OLAP is crucial; however, it is costly to support OLAP for a large data-set. An approximate query process (AQP) was proposed to efficiently compute approximate values as close as to the exact answer. Existing AQP techniques can be categorized into two parts, online aggregation, and offline synopsis generation, each having its limitations and challenges. Online aggregation-based AQP progressively generates approximate results with some error estimates (i.e., confidence interval) until the processing of all data is done. In Offline synopsis generation-based AQP, synopses are generated offline using a-priori knowledge such as query workload or data statistics. Later, OLAP queries are answered using these synopses. This paper focuses on surveying only the online aggregation-based AQP. For this purpose, firstly, we discuss the research challenges in online aggregation-based AQP and summarize existing approaches to address these challenges. In addition, we also discuss the advantages and limitations of existing online aggregation mechanisms. Lastly, we discuss some research challenges and opportunities for further advancing online aggregation research. Our goal is for people to understand the current progress in the online aggregation-based AQP area and find new insights into it.
\end{abstract}


\maketitle

\section{Introduction}
In recent world, there is an explosive growth in volume of data in various fields such as medical sectors, government sectors, education services, business and retailing sectors. This huge volume makes the big data unavoidable as it involves many individuals and spaces in daily basis. Moreover, the big data is associated with huge scientific research interests from researchers around the world because of its massive commercial value. Therefore, the value of supports for scientific decision making is recognized and also be developed around the world. Many organizations have already started mining the valuable information and knowledge from big data collected every day and have utilized the actual value by fulfilling real needs in improving their performance. As this improves efficiency and reduces the cost of industries, analyzing big data can grow enterprise innovation by creating emerging knowledge industries such as data manufacturing and services. 

One of the important type of data analysis is online analytical processing (OLAP) which is at the core of the data management and analytical systems' functionality \cite{duan2018bus}. Many applications use OLAP for making many important decisions online, such as business intelligence, in which case, the performance of OLAP is crucial. Generally, the initial stage is data analysis tasks like OLAP is to explore data by using some of the rudimentary statistics of the dataset like sum, mean, minimum/maximum, and so on. However, although these types of operations may sound very simple at first glance, the massive amount of data associated with them makes them very complex and time-consuming in real-world scenarios. Thus, to support OLAP for large dataset is costly and especially it is severe in case of big data. Many systems have been developed to handle OLAP for big data, such as Spark SQL, Pig, and Hive. For executing an OLAP query, these systems generally take approximately tens of minutes or hours for some cases. However, as many applications have the requirement of getting results from OLAP queries in seconds, these systems are practical in such cases. 

To solve the above problem, the need for approximate query processing (AQP) has appeared in existing systems. For processing extensive data, AQP efficiently approximates the original answer with some guarantee of the quality of the estimation result that fulfills the requirement of many high-performance-oriented systems. Existing AQP techniques can be broadly categorized into two types, online aggregation and offline synopsis generation. Online aggregation-based AQP progressively generates approximate results with some error estimates (i.e., confidence interval) until all data is processed. In Offline synopsis generation-based AQP, synopses are generated offline using a-priori knowledge such as query workload or data statistics. Later, OLAP queries are answered using these synopses. In this paper, we focus on surveying online aggregation-based AQP techniques. To understand online aggregation clearly, let us use an example query:
\begin{verbatim}
SELECT CourseID, AVG(Score) FROM 
Courses WHERE Score > 70 GROUP BY CourseID;
\end{verbatim}
where AVG represents an aggregate operation. One very straightforward method of online aggregation is to do random sampling to get a set of \textit{Courses} samples such as $c_1, c_5, c_{20}$. Generally, how many samples are to be selected depends on the time constraints determined by the users. Now, after getting the samples of data instances, the task to estimate the result of that query from that sample and show the error estimates (mostly as a confidence interval) to the user. Here, the goal of online aggregation is to find an approximate answer of a given query as close as to the exact answer of that query very efficiently. 

Hellerstein first proposed online aggregation in 1997 \cite{hellerstein1997online}. It was groundbreaking work in the database aggregation system. Before that, aggregation operation in the database was mainly executed in batch mode. In general, in an aggregation query, we want to get some attributes of data that require scanning the whole dataset most of the time, which is a lot of time waste. However, a user naturally does not want to spend much time getting the result from a single query. This scenario is where the motivation for online aggregation rises. Online aggregation provides users with an interactive way of dealing with aggregation queries in large datasets.  More specifically, the size of samples is incrementally updated by the system continuously, and the current result and error estimates are refined. Furthermore, users can stop the query while they think the result is sufficiently precise and proceed with the following query. This interactive way of handling aggregation queries improves efficiency by a large margin as the user does not have to spend a long waiting period for finishing the whole query.

There are several surveys conducted on approximated query processing (AQP) as such \cite{cormode2012synopses,mozafari2017approximate,chaudhuri2017approximate,kraska2017approximate,li2018approximate}. These previous surveys mainly focused on the classical aspects of AQP \cite{cormode2012synopses} like offline synopses such as Sample, Histogram, Wavelet, and Sketch without considering the online AQP scenarios where the survey paper \cite{chaudhuri2017approximate} focused on online aggregation. In \cite{kraska2017approximate}, the authors organized their survey focusing on the new challenges and opportunities such as interface, effective query planning, and database learning theories. In \cite{mozafari2017approximate}, the survey on AQP is done from the perspective of interactive data science. The most recent survey paper on AQP \cite{li2018approximate} tries to cover works before 2018 in both online and offline AQP. However, none of the above studies surveys the wide spectrum of approaches concentrating on online aggregation-based AQP. 

Recently, a survey on online aggregation has been conducted in \cite{li_2018}. However, the authors of this paper only focus on a special type of online aggregation approach that is sampling-based methods and did not review other wide range of mechanisms available for online aggregation in the existing literature. Therefore, in this paper, we survey AQP methods only focusing on online aggregation. For this purpose, we firstly discuss the research challenges in online aggregation-based AQP and summarize the existing approaches to address these challenges. Secondly, we review the advantages and limitations of existing online aggregation mechanisms. Lastly, we discuss some research challenges and opportunities for further advancing online aggregation research. Having a clear and complete understanding of approaches based on online aggression encompassing various problems regarding online aggregation will help the researchers to advance this specific area.

\begin{figure}
\centering
\centerline{\includegraphics[width=0.8\linewidth]{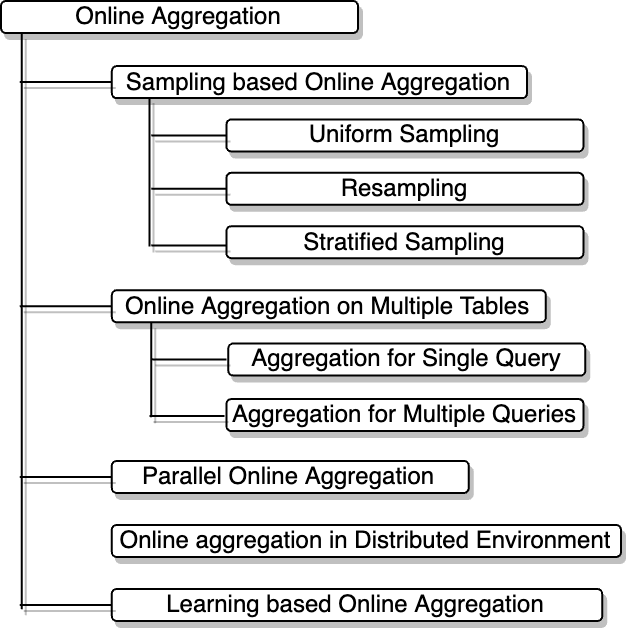}}
\caption{Taxonomy of online aggregation methods}
\label{fig:taxonomy}
\end{figure}

The rest of the paper is organized as follows. We first discuss sampling-based online aggregation methods along with error estimation mechanisms in Section \ref{sec:ola_sampling}. Section \ref{sec:ola_mult} reviews the existing methods used in online aggregation on multiple tables. In section \ref{sec:ola_parallel}, we discuss the existing approaches regarding parallel online aggregation followed by the discussion on approaches of online aggregation in distributed setting in Section \ref{sec:ola_dist}. Section \ref{sec:ola_learn} reviews research on learning-based online aggregation methods. Finally, Section \ref{sec:conclusion} concludes the survey with a summary and future research direction. The taxonomy of online aggregation reviewed in this paper is shown in Figure \ref{fig:taxonomy}.

\section{Sampling based Online Aggregation}
\label{sec:ola_sampling}
\subsection{Sampling Methods}
In sampling based online aggregation, first decision that needs to be taken is how to generate random samples for approximating the original data.
\subsubsection{Uniform Sampling}
Uniform sampling is one of the simplest way of doing online aggregation. In this technique, each tuple in the table is sampled with the same probability \cite{olken1995random,olken1986simple,chaudhuri1999random,piatetsky1984accurate}. The reason behind using uniform random sampling is that many real-world datasets have uniform or Gaussian like distributions. That's why, in early stages of sampling based online aggregation literature, the researcher was more focused on uniform sampling in a single table. Given the assumption of the uniform data distribution, random sampling approaches generate a confidence interval for the approximation of original results in most the cases. For example, \cite{hellerstein1997online} selects samples online uniformly on the fly and let users to both observing their aggregation queries' progress over time and controlling the execution. The users naturally take the decision of execution by observing the confidence interval shown for each step of online updates. To make sure that the samples are retrieved in random order, they use head scans and index scans. The success of random sampling depends on how random the samples, more random means more accurate estimations. 

Recently, sampling approaches are used in window function based online aggregation where instead of using single tuple to do aggregation, a set of tuples within a specific range is used. For instance, to handle the problem of processing big data in window functions, \cite{song2018approximate} uses some of the algorithms used in online aggregation by proposing two mechanisms: range-based global sampling algorithm and row-labeled sampling algorithm. However, as discussed before, the uniform sampling assumes the uniformity of datasets, many real world scenarios do not follow this assumption, making this approach vulnerable for those cases. On the other hand, the mechanism for sampling uniformly incurs high cost to get random samples.

\subsubsection{Resampling}
While accessing datasets multiple times for drawing random samples, it creates large I/O for the systems. To solve this problem, a sampling approach using bootstrap mechanism is used for online aggregation. More specifically, in bootstrapping, instead of accessing whole dataset, it makes a specific number of trials on a subset of datasets with replacements. Here, multiple trials (or resampling) are done for the purpose of robust estimation of standard errors and confidence intervals of the original dataset. For the case of online aggregation, however, as repeated sampling needs to be done, this creates the high computational cost for drawing samples. In \cite{agarwal2014knowing,zeng2014analytical}, the researchers solve this problem with reducing the computational cost  by using an integer value for a simulation of the probability distribution of bootstrap samples. On the other hand, a very recent resampling approach \cite{park2018verdictdb} uses variational subsampling mechanism instead of resampling same number of samples every time. Therefore, by lowering strict requirement of same size sampling, this method improves the efficiency of aggregation. 
\subsubsection{Stratified Sampling}
When there is skewness in the data, we can see the limitation of uniform sampling method. More specifically, the uniform sampling methods may not select items from a minority group with a very small number of samples. Therefore, the selected samples will lose the representativenss of whole dataset. To solve, this problem, existing approaches use stratified sampling mechanism where the items selected in sub sampling are from all representative groups of original data with same proportion. For example, in \cite{an2010dynamic}, the authors first clusters the data into non-overlapping groups and apply subsampling from each group and linearly combine the subset into one. Therefore, the proportion from each subset remains in the resultant selected subset. Another advantage of this method is that it does not assume random distribution of dataset and as it can adjust the clusters dynamically, it can highly reduce the I/O cost for sampling in online aggregations. Similarly, \cite{an2010dynamic,joshi2008robust} samples same number of tuples from each group using stratified sampling mechanism.

\subsection{Error Estimation}
\begin{figure}
\centering
\centerline{\includegraphics[width=\linewidth]{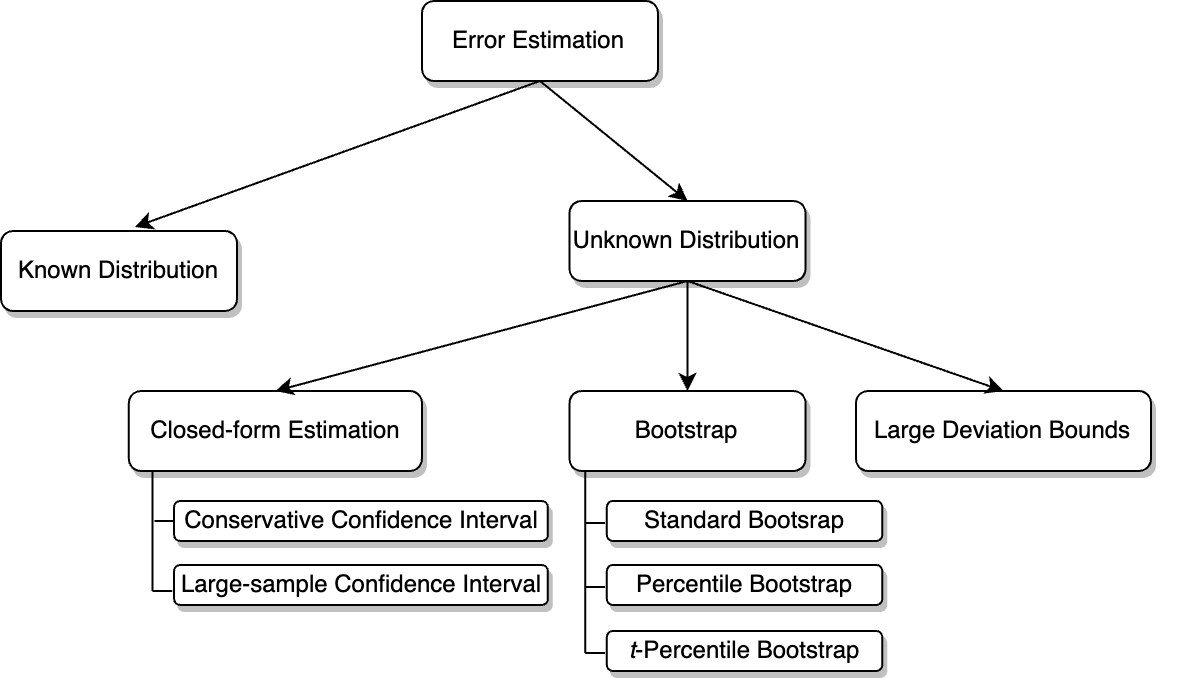}}
\caption{Overview of error estimation}
\label{fig:error}
\end{figure}

One crucial feature of random sampling approaches is to estimate the error over generated samples. One of the most widely used approaches for showing the quality of the result is to use a confidence interval. A numerical interval with a confidence value is shown to users using the statistical theory in each confidence interval.

The choice of how to estimate the error of a sample depends on whether the distribution of the data is known or unknown. We get a sample based on the sampling methods described in the previous section. If the data distribution is known prior, the sample can be used to estimate its distribution and then estimate the error based on the distribution. On the other hand, if we do not know about the distribution of the data, first, it needs to estimate the sampling data distribution and then calculate the error estimation.

In the following subsequent sections, we will discuss these two scenarios of known and unknown data distribution of error estimation in detail with corresponding methods to do the estimation. The overview of error estimation is shown in Figure \ref{fig:error}.

\subsubsection{Known Distribution}
In real-world applications, many datasets follow the normal distribution, and many studies in the literature also make an assumption that the data follows the normal distribution. If we know the distribution of the respected data before or get the distribution as the data size is large enough; then it is a classical parameter estimation problem.
Now, from known data distribution, we can take either of two approaches to calculate confidence interval based on the below two scenarios:

\subsubsection*{Known Variance}
When the variance of the distribution is known, the confidence interval can be easily calculated using Gaussian distribution $\mathcal{N}\{\mu, \sigma^2\}$

\subsubsection*{Unknown Variance}.
On the other hand, if the variance of the distribution is not known, we can assume this as a $t$-distribution and formalize the confidence interval using an approach illustrated in \cite{li_2018}. In such an approach, as the sample size increases, we can get higher confidence and lower interval as it provides more information about the actual population.

\subsubsection{Unknown Distribution}
According to \cite{li_2018}, for the case of unknown data distribution, calculation of confidence interval can be usually done using three methods. They are Closed-form Estimation, Bootstrap and Large Deviation Bounds.

\subsubsection*{Closed-form Estimation}

This approach ensures that the results from samples fall into a confidence interval with a specific probability. For instance, consider that $n$ and $R_s$ are the numbers of samples ($s$) and the sample's aggregate result, respectively, and $R$ is the actual result. In advance, we assume a value for the confidence $p$ such as 95\%. Then the goal is to calculate a variable $\eta_s$ to get the confidence interval $[R_s - \eta_s, R_s+\eta_s]$. According to \cite{hellerstein1997online}, the confidence interval in closed-form can be categorized into two categories: conservative confidence interval and large-sample confidence interval.

\begin{enumerate}
    \item {\textit{Conservative Confidence Interval.} To obtain a conservative confidence interval, the Hoffing inequality is used, which ensures that the probability that $R$ falls into $[R_s - \eta_s, R_s+\eta_s]$ is greater than or equal $p$. The Hoffing inequality is:
    \begin{equation}
        p(|R_s - R| \leq p) \geq 1 - 2e^{-2n\eta_s^2/(b-a)^2}
    \end{equation}
    where there exists constants a and b, known a priori \cite{hellerstein1997online}. We can get the value of $\eta_s$ by setting the right side of the inequality to $p$. The conservative confidence interval usually is not used in practice. Because it can only be applied in a very loose condition to achieve that is when n is 1.
    }
    \item {\textit{Large-sample Confidence Interval.}
    To obtain the large-sample confidence interval, the central limit theory is used. It ensures the probability of $R$ falling into $[R_s - \eta_s, R_s+\eta_s]$ is equal to $p$. When $n$ is sufficiently large, the mean of sample follows a normal distribution with mean $\mu$ and variance $\sigma^2$, i.e. $R_s \sim \mathcal{N}(\mu, \sigma^2)$. Here the $\mu$ and $\sigma^2$ are the mean and variance of the original data. As the value of the variance $\sigma^2$ is not known for the original data, here the variance of the samples is used instead that is $T_{n,2}(v) = \sum_{i=1}^n (v(i) - R_s)^2$ \cite{hellerstein1997online}. Thus, we can write,
    
    \begin{align}
        p(|R_s-R| \leq \eta_s) = p(|\frac{\sqrt{n}(R_s - R)}{T^{1/2}_{n,2}(v)}| \leq \frac{\eta_s\sqrt{n}}{T^{1/2}_{n,2}(v)}) \nonumber\\ \approx 2\phi \big( \frac{\eta_s\sqrt{n}}{T^{1/2}_{n,2}(v)} \big) - 1
    \end{align}
    where, by setting the equality to 1, we can get the value of $\eta_s$. When the number of samples is large enough to follow the normal distribution, the CLT can be used to calculate the confidence interval. We get a tighter interval from a large sample confidence interval than the conservative one. Since it does not require the original data to follow any distribution, it is often used in online aggregation.
    }
\end{enumerate}

\subsubsection*{Bootstrap.}
The above-specified estimation method does not work when the number of the samples is very small as it provides a very loose confidence interval. As said before, since the bootstrap \cite{efron1992bootstrap} resamples on the samples, its accuracy does not depend on the sample size. In the following sections, we discuss the three most popular bootstrap methods: standard bootstrap, percentile
bootstrap and t-percentile bootstrap \cite{efron1992bootstrap}.

\begin{enumerate}
    \item \textit{Standard Bootstrap.}
    The standard bootstrap resamples the original data $k$ times of size $n$ with replacement.  From $k$ resamples, therefore, $k$ samples $\{s_1,\cdots, s_k\}$ is found and for each resample, an estimation of true result can be get. Let $\tilde{s}$ be the result estimation of $s$ and $\tilde{s}_i$ be the result estimation based on the i-th resample $s_i$, then:

    \begin{align}
        \tilde{s} = \frac{1}{k} \sum_{i=1}^k\tilde{s}_i, \quad \sigma^2_{\tilde{s}} = \frac{1}{k-1} \sum_{i=1}^k (\tilde{s}_i - \tilde{s})
    \end{align}
    where $\sigma^2_{\tilde{s}}$ is the variance of $\tilde{s}$. Therefore, with $(1-\alpha)\%$ confidence interval, the true result lies in:
    \begin{align}
        [\tilde{s} - \mu_{1-\alpha/2} \sigma^2_{\tilde{s}}, \tilde{s} + \mu_{1-\alpha/2} \sigma^2_{\tilde{s}}]
    \end{align}
    \item \textit{Percentile Bootstrap.}
 The upper bound and lower bound of the statistical confidence interval with the confidence of $(1-\alpha)$ is equal to the $\alpha/2$ th quantile and $(1-\alpha/2)$ th quantile of the empirical distribution of the bootstrap \cite{efron1992bootstrap}. It is the main principle used in the percentile bootstrap to sort the $k$ bootstrap resamples estimate $\tilde{s}$. If the sorted $\tilde{s}$ is denoted as $\tilde{s}^*$ then the true results lies in:
    
    \begin{align}
        [\tilde{s}^{*}_{\frac{\alpha}{2}k}, \tilde{s}^{*}_{(1- \frac{\alpha}{2})k}]
    \end{align}
    
    \item \textit{$t$-Percentile Bootstrap.} 
    t-Percentile bootstrap is a modification of percentile bootstrap. It calculates $t$ statistics of each resample estimate $\tilde{s}_i$ i.e., $t(i) = \frac{\tilde{s} - \tilde{s}_i}{\sigma_{\tilde{s}}}$. 
    Let $\alpha$ be the significance level, the $\alpha/2$ th quantile and $1- \alpha/2$ th quantile are $t^*_{\frac{\alpha}{2}k}$ and $t^*_{(1-\frac{\alpha}{2})k}$ respectively. Thus, the true result lies into the confidence interval with probability $(1-\alpha)$:
    
    \begin{align}
        [\tilde{s} - t^{*}_{\frac{\alpha}{2}k}\sigma_{\tilde{s}}, \tilde{s} + t^{*}_{(1- \frac{\alpha}{2})k}\sigma_{\tilde{s}}]
    \end{align}
\end{enumerate}

\subsubsection*{Large Deviation Bounds}
A large deviation bound \cite{mcdiarmid1998concentration} is calculated by considering the sensitivity of outliers. Thus, it is considered as the worst-case version of the confidence interval. The confidence interval by the large deviation bound depends on the corresponding aggregation function. For instance, if the aggregation function is SUM, then MAX and MIN are used as the bound for the confidence interval. Generally, it computes an interval that is much wider than the real width of the sample distribution.

\section{Online Aggregation on Multiple Tables}
\label{sec:ola_mult}

Generally, queries on a single table are simple and account for a small proportion of the entire query operations. However, complex OLAP queries associated with multiple tables are pretty common in real-world applications. These types of queries naturally contain join-predicates. 

For example, the following query is for getting the lowest User age from Users table for each CityName in the Cities table.
\begin{verbatim}
SELECT Cities.CityName, MIN(Users.Age)
FROM 
Cities JOIN Users ON Cities.ID = Users.cityID
GROUP BY Cities.CityName
\end{verbatim}

Here, to get the result for this specific query, we need to join two tables based on their common attributes (i.e., cityID). However, in the database system, the join query is regarded as one of the most expensive queries to be processed. More specifically, during the execution of entire query processing, joins cost the most of time or space. This thing gets worse when the dataset is too big. Sample problem happens in case of online aggregation associated with join queries of multiple tables. In real-world applications, the data size is generally large, and thus, for join query, there needs to do millions of computations for join operation.

\subsection{Aggregation for Single Query}

In literature, at first, the online aggregation was designed for only a single table. One straightforward way to join multiple tables in the online aggregation setting is to join all the tables. Then, apply online aggregation approaches on the already joined tables to estimate the result. However, it is easy to say that this method will incur high costs on joining tables.

To address the above problem, Hass extends single table online aggregation to adapt it in case of multiple tables, and the approach is called ripple joint \cite{haas1999ripple,haas1997large,jermaine2008scalable}. In this proposed approach, the authors consider joining multiple tables using nested-loop ann hash join. The purpose of this approach is to get as close estimation as possible efficiently to obtain satisfactory accuracy. Specifically, the approach is based on the sampling mechanism. The main idea behind the approach is to generate samples from each table and use these samples to join the tables iteratively as long as the confidence interval is not satisfactory. As it samples continuously, there is no problem of non-uniform distribution that appears in traditional join query with two uniform random samples. 

However, one problem with the above approach is that many samples cannot be joined, and thus some groups will be left out in the result. Another problem with ripple join is that it performs poorly with memory overflow issues. To solve this problem, Luo et al. \cite{luo2002scalable} modify ripple join for multiple tables using simultaneous sampling with the execution of merge-sort join. Therefore, it does not suffer from poor performance in the case of memory overflows. Several works \cite{dittrich2002progressive,jermaine2006sort,jermaine2005disk} have been done to improve conventional join algorithms based on the idea of ripple join. However, as the main idea is very close to ripple join, these approaches also suffer from the limitations of ripple join. For example,  these approaches are expensive and assume that stored data is randomly distributed. Therefore, these approaches are not adaptive to query optimization algorithms like sorting and indexing as they do not make such assumptions. 

To solve the limitations of ripple join, a recent approach called "wander join" is proposed in \cite{li2016wander0,li2016wander,li2019wander}.
This algorithm uses the idea of random walks over the join graph constructed by considering a tuple as a node in the graph, and there is an edge between two nodes if their corresponding tuples can join. It does not make any assumptions about the data. Furthermore, unlike ripple join, which blindly selects samples, "wander join" selects the samples that would be joined in every next step. The authors show that wander performs better than ripple join in both cases, whether memory overflows or not.

From the above discussion, we observe that if it is possible to know the join size in advance, it is easy to make a better join plan. However, computing exact join size is rather costly, and thus we need to estimate the size of the join. To estimate the size of a join, the existing approaches mainly join the synopses of each table rather than scanning the whole dataset. For instance, a recent study proposes a new algorithm to estimate the size of equijoin over multiple tables \cite{vengerov2015join}. 
The proposed correlated sampling approach constructs a small space synopsis for each table. This synopsis is then used to calculate the estimate of the join size of the specific table with others. This algorithm is suitable for streaming applications as it only makes a single pass over the data.

\subsection{Aggregation for Multiple Queries}
An interesting observation from the above discussion of online aggregation over multiple tables is that the above studies only work for a single query and do not share supporting between multiple queries. It is a serious limitation restricting the performance of online aggregation. As the queries are processed independently without considering their interactions of sharing opportunities, it incurs two additional costs of execution: (1) the high redundant I/O cost and (2) the cost for replicative statistical computation. To solve this problem, Wu et al. first proposed an algorithm \cite{wu2010continuous} where a Dynamic Acyclic Graph (DAG) is used to organize multiple queries such that the interaction between multiple query results can be realized. By taking advantage of intermediate results, the authors achieved better efficiency.

On the other hand, in \cite{wang2014oats}, the authors proposed an online aggregation with a two-level sharing strategy in the cloud (OATS). This approach is built upon the architecture of Map-Reduce and effectively supports online aggregation in large-scale concurrent query processing on skewed data distribution. In the first level of sharing, this method utilizes a sample buffer strategy for storing sampling results such that the sharing among multiple online aggregations queries can be done, thus reducing the additional computational cost. The algorithm shares partial statistical computations to reduce the replicative statistical computation cost, reducing the number of operations needed for final aggregations. However, although sharing helps to reduce the computation cost or repetitive sampling, it requires extra storage for intermediate results. Therefore, this is a trade-off between cost and space.

\section{Parallel Online Aggregation}
\label{sec:ola_parallel}
With advent of new hardware machines, many time-consuming tasks can be done using the power parallel computing. Therefore, it has improved the efficiency of many large programs in computing. Similarly, many mechanism are also proposed in online query processing to accelerate its execution. Because of these parallel techniques, wasting extra time for estimating error during query processing has been be avoided. 

As an early work, in \cite{luo2002scalable}, the authors propose an approach that extends the centralised ripple join algorithm \cite{haas1999ripple} to a parallel settings. In this paper, the proposed algorithm named parallel hash ripple join combines sampling with parallelism using a non-blocking hybrid parallel hash join algorithm. For the sampling, this algorithm makes use of a stratified sampling estimator \cite{singh2018sampling} to estimate the result while it is not always possible to derive confidence interval. The authors shows that their method maintains good performance in memory overflow scenario. In \cite{wu2009distributed}, a parallel online aggregation algorithm was proposed by extending online aggregation to distributed point-to-point (P2P) networks. For this purpose, a synchronised sampling estimator was introduced where the algorithm requires to move data from storage nodes to processing nodes while executing the system. However, the problem with this approach is that it shows poor performance in case of highly parallel systems that are not synchronous. In a later work \cite{wu2010continuous}, the authors design online aggregation algorithm that works over multiple queries by utilizing the benefits of global random sampling method.

In recent times, there is a growing interest in research on the area of online aggregation in Map-Reduce. The main reason for this is that the original Map-Reduce implementation in Hadoop \footnote{http://hadoop.apache.org/} performs poorly which makes the use of online aggregation in this context a crucial need, specially for big data processing. For this purpose, the first attempt was to extend Hadoop framework in \cite{condie2010mapreduce}. The authors in this paper proposed an algorithm called Hadoop Online Prototype (HOP) that facilities parallel execution of Hadoop for online aggregation by allowing extraction of partial aggregates during execution. This was possible by using pipelining between operators. Moreover, this Hadoop implementation also supports continuous queries, for which, the applications like  event monitoring and stream processing are possible by writing MapReduce programs. By enabling online aggregation, the HOP retains the fault tolerance properties of original Hadoop containing the blocking operators and thus is possible to run unmodified user-defined MapReduce programs. 

HOP enables Online aggregation queries by executing tasks reducing data in different intervals e,g., $10\%,20\%, \cdots, 90\%$. The query estimation in Hadoop assumed a uniform sample of the input data. However, it made no modifications for enforcing this property in the Hadoop scheduler program. That's why, this led to a significant error in while making the estimations. To solve this problem, the authors made a change to the query to scale the estimate accordingly to a extra parameter denoting the number of samples of particular group of aggregates present in the query. However, did not use any formal sampling or estimation method, rather than doing only the partial result extraction. Therefore, HOP is actually regarded as a partial aggregation system rather than an online aggregation system. 

The first work that completely transforms Map-Reduce for a real online aggregation system is \cite{pansare2011online}. In this paper, the author considers building online aggregation into a Map-Reduce system concerned for large-scale data processing. To handle the correlation between the data partition process time and result generation time, the authors utilize a Bayesian framework to model their algorithm. The reason for using this type of model stems from the nature of chunks used for producing an aggregate. As the chunks do not carry any information with them about their content or the operation being performed in the aggregate generation, it is treated as black boxes. The proposed model makes a prediction for the aggregates being performed on the yet non-scheduled chunks based on the chunks that have already been processed and the time to schedule and process the chunks. Here, the processed chunks are distributed from the whole chunk collection independently and identically. The proposed Bayesian model is dynamic in the sense that it keeps itself updated continuously as more chunks are processed, which allows it to estimate more accurately as more data are being processed.

BlBlinkDB [4] is a sampling-based parallel online aggregation method for executing interactive queries on large data. For facilitate parallel computing, this approach makes different sized pre-computed samples and store them on disk. However, this makes a problem on massive dataset. Because for storing pre-computed data, along side original data, the systems requires to store huge amount of additional data. Furthermore, pre-computation of sample of different sizes waste logn period of time even if it is done offline. Like iterative online aggregation, a query is evaluated on a selected sample, producing an estimate. The accuracy on this esitmate is judged and if it is not satisfying, the execution of next query on large data sample is done. By doing such way, there maybe case where the query need to be executed on the whole dataset. Therefore, although it can allow growing accuracy on discrete estimates, this is not possible for continuous estimation. A similar approach to BlinkDB is proposed in \cite{laptev2012early} named EARL. Rather than pre-computing a large number of samples, the EARL first pre-computes a single sample and then uses bootstrapping to make multiple samples from the pre-computed single sample. Then, these multiple samples are used to compute estimates. This process is continued dynamically at runtime, improving the accuracy with increased samples.

To support the continuous estimates and confidence intervals in a distributed system for online aggregation, Qin and Rusu first proposed a generic framework \cite{qin2013parallel}. This approach enables the execution of very complex queries associated with a massive amount of data. Moreover, this approach is also the first to support multiple estimators when there are different online aggregation strategies. In a later work, Qin and Rasu proposed a common framework called PF-OLA \cite{qin2014pf} which can model a much larger class of estimation methods. This framework aims to execute online aggregation parallelly where the estimated results and their corresponding confidence interval are continually refined. Here, the refinement process is based on samples selected during query processing. This is also the first work to do parallel online aggregation where except for the actual execution, there is practically no overhead of execution. 

For dealing with the problem of arbitrarily nested aggregates, an online aggregation-based architecture named G-OLA \cite{zeng2015g} is proposed that uses efficient delta maintenance techniques. More specifically, G-OLA performs random partitions of the whole datasets to construct multiple smaller batches and, at a time, works with a single mini-batch to process the whole data. It then partitions intermediate results into uncertain and deterministic sets and performs updates in new mini-batch while dealing with uncertain sets.

Unlike the above approaches, OLA-RAW is a bi-level sampling method designed as a parallel online aggregation used in raw data \cite{cheng2017bi}. Sampling in OLA-RAW is query-based and performed exclusively in-place when the query is at run time, without reorganizing the data. This is accomplished by a new resource-aware bi-level sampling mechanism that simultaneously processes data in random blocks and adaptably determines how many tuples need to be sampled in a block. OLA-RAW avoids the high cost incurred for repetitive conversion from raw data by incrementally creating and maintaining a bi-level sample synopsis residing in memory.

\section{Online aggregation in Distributed Environment}
\label{sec:ola_dist}

Because of the massive data size, many real-world applications use distributed clusters to store their data. Therefore, while a complex join operation is associated with multiple tables stored in multiple clusters, the data need to be scanned repetitively. More specifically, it will create high communication costs during join operations as the data must be transmitted from one cluster to another. To illustrate this scenario, let us discuss an example where a business analyst wants to know the most popular brands in super shops being sold during holiday seasons. To answer this query, the systems need to scan multiple tables regarding shops, orders, and customers where these tables are distributed in many clusters. Therefore, there is a computational overhead because of multiple scans from different clusters to generate the result of a join query.

To facilitate online aggregation with join operations in distributed settings, the existing approaches try to overcome two challenges: (1) avoid multiple scanning of data for reducing I/O cost and (2) reduce the communication cost among distributed clusters. In the early research stage regarding this matter, the authors try to discover efficient sampling methods that can be appropriate for a distributed environment. For example, in \cite{haas2004bi} proposed a bi-level sampling mechanism to combine the methods of row-level and page-level samplings used in distributed systems. However, these types of approaches are shown to be inefficient in more complex query predicates.

To solve the above problem, in one later study \cite{laptev2012early}, an algorithm called Early Accurate Result Library (EARL) is proposed where a system was designed to reduce the changes required in the Map-Reduce environment. This method is a non-parametric extension of Hadoop, which allows two items: (1) it permits to do the cumulative computation of early results generated from arbitrary workflows and (2) the degree of accuracy that has been achieved in the incremental process can be reliably estimated. For this purpose, the proposed approach takes leverage of bootstrapping to generate online uniform samples and then evaluates the accuracy so far from the samples accumulated incrementally. Another advantage of this approach is that it can handle complex queries well. However, one limitation of this method, along with a similar method called ApproxHadoop \cite{goiri2015approxhadoop} is that it assumes the dataset clustered in different nodes are uniformly distributed. They do not consider the case of skewness in the dataset.  Therefore, if the cluster dataset stored in each node is not uniformly distributed, they cannot operate well by generating biased sampling towards one group of samples.

In a later approach named Sapprox \cite{zhang2016sapprox}, the limitations of the above approaches were overcome. To do so, Sapprox constructs a probabilistic map called Segmap offline to capture the skewed data distribution in subsets of data clustered in different nodes. Then, this information of skewed distribution was informed to a proposed online sampling mechanism to efficiently sample data for each subset in the distributed cluster environment. Therefore, the final sampled data does not suffer from a biased estimation of the original dataset. Moreover, for distributed file systems, this method considers quantifying the optimal size of the sampling unit before doing sampling for generating estimation.

To improve performance of online aggregation in distributed cluster environment, Kandula et al. proposed a new approach called Quickr \cite{kandula2016quickr,kandula2016errata} which can select an appropriate sampling plan to draw samples for each cluster. For this purpose, the proposed framework takes leverage of the advantages of three sampling operators by combining them. These three sampling operators are uniform sampling method, distinct sampling method, and universal sampling method. In distinct sampling, samples for each group are selected, and samples for generating join results over multiple tables are selected in the universal sampling method. During the first pass of the sampling, it uses the lazy sampling approach to efficiently scan the datasets distributed over all the clusters. For calculating error estimation of the samples, it uses different strategies concerns for different aggregate types. For instance, on the one hand, when the aggregation and group by operator follow immediately after a sampler operator, for calculating the confidence interval, it uses the central limit theorem by extending the well-known Horvitz–Thompson (HT) to estimate the actual answer. On the other hand, it utilizes the theory of “dominance transitivity”.

\section{Learning based Online Aggregation}
\label{sec:ola_learning}

Traditionally, in database systems, the vast amount of previous query answers are not naturally used to process future queries.
With the advent of modern machine learning (ML) approaches (i.e., deep learning models), if we can use previous queries to predict future queries, it can also be a good option for answering new queries. It is the primary motivation behind a recent approach called "Database Learning" (DBL) \cite{park2017database}.

Motivated from the machine learning principle where we can learn a model to predict future events based on past events, DBL proposed to learn to answer new queries based on past queries and their corresponding answers. To be more specific, as the training set for the ML model, this approach considers observations of past queries (with results) and uses this for calculating posterior knowledge regarding a target dataset for the later purpose of fastening the future queries. To train a model for representing the distribution inherent in the data, DBL uses statistical features such as correlation and covariances between all pairs of query snippets from the past query collections. It is not easy to know the whole data distributions a sample belongs to whenever it is used to answer the new queries. Nevertheless, this problem is solved by using the answers related to the past queries. Because, when we can estimate the distribution of the data based on the past data and use that to infer the result for new query based on the model learned using statistical features and known labels (i.e., query answers).

Therefore, we can learn a more precise model using more previous data, which means the system requires less number for actual data (a smaller sample) to answer the query. As the required sample size is small, it can faster response time. However, the success of this learning-based approach strongly depends on the quality of the training set. If the past query results used to learn the model are not accurate, then the learned model also performs poorly to infer future query results. Therefore, it may continue to make false inferences as long as the quality of the past data deteriorates. 

Although not an utterly learning-based approach, the algorithm proposed in \cite{galakatos2017revisiting} called incremental AQP, uses quite similar motivation for online AQP. This method introduces a new formulation for AQP, considering aggregate query results as random variables. By doing so, it facilitates reusing approximate query results that reason about the error propagation on overlapping queries. For a new query, it first gets the past queries with shared attributes and query conditions, then uses the results of those past queries to refine the approximation. This work supports the queries interactive to rare groups by using a probability model with the help of a low-overhead partial index and corresponding rewriting rules.

Different from DBL, which learns from queries, an approach called DeepDB \cite{hilprecht2019deepdb} was proposed as a data-driven method (not from queries) for the database components learning. To capture the critical properties of a database, this work proposed deep probabilistic models over databases called Relational Sum-Product Networks (RSPNs). The author identified that this model is generally applicable in applications like cardinality estimation for a query optimizer and approximate query processing (i.e., online aggregation). From the perspective of online aggregation-based AQP, this method supports a faster response time of answering a query in a large dataset by allowing aggregate queries with equi-joins and selection predicates with group-by clauses.

Another work that does not use the queries to learn the model is \cite{kulessa2018model}. In this work, the main is to learn a generated model over the original potentially large dataset and then use this model to answer new SQL queries interactively for data exploration. 
A sampling-based online AQP using a deep generative model was proposed in \cite{thirumuruganathan2020approximate}. This method learns data distribution using a deep generative model for generating samples for the AQP. More specifically, the authors of this paper explore the use of DL methods to answer aggregated queries in data visualization and exploration. It demonstrates that DL's effective and efficient manners can be used for solving AQP problems for higher performance in this task.

\section{Conclusion and Future Direction}
\label{sec:conclusion}
This paper reviews extensive studies on online aggregation-based approximate query processing. It has been more than twenty years since the first research about online aggregation. However, the interest in this research topic has not decreased, and it is still in the academic research stage. This paper first summarizes the existing online aggregation methods in different scenarios by adequately designing a research taxonomy on this field. We discuss challenges and summarize the mechanisms used in the literature. We also discuss the advantages and limitations of the existing approaches. In summary, we categorize the online aggregation methods into five broad categories based on their solution and application characteristics.   
By reviewing the existing approaches, we found out that improving the following aspects can be new research directions in this field.

\begin{itemize}
    \item Existing online aggregation methods assume a distribution for the dataset or compute the distribution from the dataset. However, if the dataset is too large to compute the data distribution, the existing will work as it depends on the knowledge of data distribution. Therefore, there needs to conduct new research regarding online aggregation on non-normal distribution data. 

    \item For learning-based online AQP, we found out that there are two lines of work: one is data-driven, and another is query-driven. Each of these two approaches has its own advantages and limitations. Therefore, combining the best of these two approaches can be an interesting future research direction. 
    
    \item The use of state-of-the-art machine learning approaches, specifically the deep learning models, is limited in existing online aggregation methods. Exploring these new mechanisms can be a new research direction on online aggregation.
\end{itemize}


\bibliographystyle{ACM-Reference-Format}
\bibliography{sample-base}


\begin{thebibliography}{52}


\ifx \showCODEN    \undefined \def \showCODEN     #1{\unskip}     \fi
\ifx \showDOI      \undefined \def \showDOI       #1{#1}\fi
\ifx \showISBNx    \undefined \def \showISBNx     #1{\unskip}     \fi
\ifx \showISBNxiii \undefined \def \showISBNxiii  #1{\unskip}     \fi
\ifx \showISSN     \undefined \def \showISSN      #1{\unskip}     \fi
\ifx \showLCCN     \undefined \def \showLCCN      #1{\unskip}     \fi
\ifx \shownote     \undefined \def \shownote      #1{#1}          \fi
\ifx \showarticletitle \undefined \def \showarticletitle #1{#1}   \fi
\ifx \showURL      \undefined \def \showURL       {\relax}        \fi
\providecommand\bibfield[2]{#2}
\providecommand\bibinfo[2]{#2}
\providecommand\natexlab[1]{#1}
\providecommand\showeprint[2][]{arXiv:#2}

\bibitem[Agarwal et~al\mbox{.}(2014)]%
        {agarwal2014knowing}
\bibfield{author}{\bibinfo{person}{Sameer Agarwal}, \bibinfo{person}{Henry
  Milner}, \bibinfo{person}{Ariel Kleiner}, \bibinfo{person}{Ameet Talwalkar},
  \bibinfo{person}{Michael Jordan}, \bibinfo{person}{Samuel Madden},
  \bibinfo{person}{Barzan Mozafari}, {and} \bibinfo{person}{Ion Stoica}.}
  \bibinfo{year}{2014}\natexlab{}.
\newblock \showarticletitle{Knowing when you're wrong: building fast and
  reliable approximate query processing systems}. In
  \bibinfo{booktitle}{\emph{Proceedings of the 2014 ACM SIGMOD international
  conference on Management of data}}. \bibinfo{pages}{481--492}.
\newblock


\bibitem[An et~al\mbox{.}(2010)]%
        {an2010dynamic}
\bibfield{author}{\bibinfo{person}{Mingyuan An}, \bibinfo{person}{Xiuming Sun},
  {and} \bibinfo{person}{Ninghui Sun}.} \bibinfo{year}{2010}\natexlab{}.
\newblock \showarticletitle{Dynamic Data-Partitioned Online Aggregation}.
\newblock \bibinfo{journal}{\emph{Journal of Computer Research and
  Development}} \bibinfo{volume}{47}, \bibinfo{number}{11}
  (\bibinfo{year}{2010}), \bibinfo{pages}{1928}.
\newblock


\bibitem[Chaudhuri et~al\mbox{.}(2017)]%
        {chaudhuri2017approximate}
\bibfield{author}{\bibinfo{person}{Surajit Chaudhuri}, \bibinfo{person}{Bolin
  Ding}, {and} \bibinfo{person}{Srikanth Kandula}.}
  \bibinfo{year}{2017}\natexlab{}.
\newblock \showarticletitle{Approximate query processing: No silver bullet}. In
  \bibinfo{booktitle}{\emph{Proceedings of the 2017 ACM International
  Conference on Management of Data}}. \bibinfo{pages}{511--519}.
\newblock


\bibitem[Chaudhuri et~al\mbox{.}(1999)]%
        {chaudhuri1999random}
\bibfield{author}{\bibinfo{person}{Surajit Chaudhuri}, \bibinfo{person}{Rajeev
  Motwani}, {and} \bibinfo{person}{Vivek Narasayya}.}
  \bibinfo{year}{1999}\natexlab{}.
\newblock \showarticletitle{On random sampling over joins}.
\newblock \bibinfo{journal}{\emph{ACM SIGMOD Record}} \bibinfo{volume}{28},
  \bibinfo{number}{2} (\bibinfo{year}{1999}), \bibinfo{pages}{263--274}.
\newblock


\bibitem[Cheng et~al\mbox{.}(2017)]%
        {cheng2017bi}
\bibfield{author}{\bibinfo{person}{Yu Cheng}, \bibinfo{person}{Weijie Zhao},
  {and} \bibinfo{person}{Florin Rusu}.} \bibinfo{year}{2017}\natexlab{}.
\newblock \showarticletitle{Bi-level online aggregation on raw data}. In
  \bibinfo{booktitle}{\emph{Proceedings of the 29th International Conference on
  Scientific and Statistical Database Management}}. \bibinfo{pages}{1--12}.
\newblock


\bibitem[Condie et~al\mbox{.}(2010)]%
        {condie2010mapreduce}
\bibfield{author}{\bibinfo{person}{Tyson Condie}, \bibinfo{person}{Neil
  Conway}, \bibinfo{person}{Peter Alvaro}, \bibinfo{person}{Joseph~M
  Hellerstein}, \bibinfo{person}{Khaled Elmeleegy}, {and}
  \bibinfo{person}{Russell Sears}.} \bibinfo{year}{2010}\natexlab{}.
\newblock \showarticletitle{MapReduce online.}. In
  \bibinfo{booktitle}{\emph{Nsdi}}, Vol.~\bibinfo{volume}{10}.
  \bibinfo{pages}{20}.
\newblock


\bibitem[Cormode et~al\mbox{.}(2012)]%
        {cormode2012synopses}
\bibfield{author}{\bibinfo{person}{Graham Cormode}, \bibinfo{person}{Minos
  Garofalakis}, \bibinfo{person}{Peter~J Haas}, {and} \bibinfo{person}{Chris
  Jermaine}.} \bibinfo{year}{2012}\natexlab{}.
\newblock \showarticletitle{Synopses for massive data: Samples, histograms,
  wavelets, sketches}.
\newblock \bibinfo{journal}{\emph{Foundations and Trends in Databases}}
  \bibinfo{volume}{4}, \bibinfo{number}{1--3} (\bibinfo{year}{2012}),
  \bibinfo{pages}{1--294}.
\newblock


\bibitem[Dittrich et~al\mbox{.}(2002)]%
        {dittrich2002progressive}
\bibfield{author}{\bibinfo{person}{Jens-Peter Dittrich},
  \bibinfo{person}{Bernhard Seeger}, \bibinfo{person}{David~Scot Taylor}, {and}
  \bibinfo{person}{Peter Widmayer}.} \bibinfo{year}{2002}\natexlab{}.
\newblock \showarticletitle{Progressive merge join: A generic and non-blocking
  sort-based join algorithm}. In \bibinfo{booktitle}{\emph{VLDB'02: Proceedings
  of the 28th International Conference on Very Large Databases}}. Elsevier,
  \bibinfo{pages}{299--310}.
\newblock


\bibitem[Duan et~al\mbox{.}(2018)]%
        {duan2018bus}
\bibfield{author}{\bibinfo{person}{Lei Duan}, \bibinfo{person}{Tinghai Pang},
  \bibinfo{person}{Jyrki Nummenmaa}, \bibinfo{person}{Jie Zuo},
  \bibinfo{person}{Peng Zhang}, {and} \bibinfo{person}{Changjie Tang}.}
  \bibinfo{year}{2018}\natexlab{}.
\newblock \showarticletitle{Bus-olap: A data management model for non-on-time
  events query over bus journey data}.
\newblock \bibinfo{journal}{\emph{Data Science and Engineering}}
  \bibinfo{volume}{3}, \bibinfo{number}{1} (\bibinfo{year}{2018}),
  \bibinfo{pages}{52--67}.
\newblock


\bibitem[Efron(1992)]%
        {efron1992bootstrap}
\bibfield{author}{\bibinfo{person}{Bradley Efron}.}
  \bibinfo{year}{1992}\natexlab{}.
\newblock \showarticletitle{Bootstrap methods: another look at the jackknife}.
\newblock In \bibinfo{booktitle}{\emph{Breakthroughs in statistics}}.
  \bibinfo{publisher}{Springer}, \bibinfo{pages}{569--593}.
\newblock


\bibitem[Galakatos et~al\mbox{.}(2017)]%
        {galakatos2017revisiting}
\bibfield{author}{\bibinfo{person}{Alex Galakatos}, \bibinfo{person}{Andrew
  Crotty}, \bibinfo{person}{Emanuel Zgraggen}, \bibinfo{person}{Carsten
  Binnig}, {and} \bibinfo{person}{Tim Kraska}.}
  \bibinfo{year}{2017}\natexlab{}.
\newblock \showarticletitle{Revisiting reuse for approximate query processing}.
\newblock \bibinfo{journal}{\emph{Proceedings of the VLDB Endowment}}
  \bibinfo{volume}{10}, \bibinfo{number}{10} (\bibinfo{year}{2017}),
  \bibinfo{pages}{1142--1153}.
\newblock


\bibitem[Goiri et~al\mbox{.}(2015)]%
        {goiri2015approxhadoop}
\bibfield{author}{\bibinfo{person}{Inigo Goiri}, \bibinfo{person}{Ricardo
  Bianchini}, \bibinfo{person}{Santosh Nagarakatte}, {and}
  \bibinfo{person}{Thu~D Nguyen}.} \bibinfo{year}{2015}\natexlab{}.
\newblock \showarticletitle{Approxhadoop: Bringing approximations to mapreduce
  frameworks}. In \bibinfo{booktitle}{\emph{Proceedings of the Twentieth
  International Conference on Architectural Support for Programming Languages
  and Operating Systems}}. \bibinfo{pages}{383--397}.
\newblock


\bibitem[Haas(1997)]%
        {haas1997large}
\bibfield{author}{\bibinfo{person}{Peter~J Haas}.}
  \bibinfo{year}{1997}\natexlab{}.
\newblock \showarticletitle{Large-sample and deterministic confidence intervals
  for online aggregation}. In \bibinfo{booktitle}{\emph{Proceedings. Ninth
  International Conference on Scientific and Statistical Database Management
  (Cat. No. 97TB100150)}}. IEEE, \bibinfo{pages}{51--62}.
\newblock


\bibitem[Haas and Hellerstein(1999)]%
        {haas1999ripple}
\bibfield{author}{\bibinfo{person}{Peter~J Haas} {and}
  \bibinfo{person}{Joseph~M Hellerstein}.} \bibinfo{year}{1999}\natexlab{}.
\newblock \showarticletitle{Ripple joins for online aggregation}.
\newblock \bibinfo{journal}{\emph{ACM SIGMOD Record}} \bibinfo{volume}{28},
  \bibinfo{number}{2} (\bibinfo{year}{1999}), \bibinfo{pages}{287--298}.
\newblock


\bibitem[Haas and K{\"o}nig(2004)]%
        {haas2004bi}
\bibfield{author}{\bibinfo{person}{Peter~J Haas} {and}
  \bibinfo{person}{Christian K{\"o}nig}.} \bibinfo{year}{2004}\natexlab{}.
\newblock \showarticletitle{A bi-level bernoulli scheme for database sampling}.
  In \bibinfo{booktitle}{\emph{Proceedings of the 2004 ACM SIGMOD international
  conference on Management of data}}. \bibinfo{pages}{275--286}.
\newblock


\bibitem[Hellerstein et~al\mbox{.}(1997)]%
        {hellerstein1997online}
\bibfield{author}{\bibinfo{person}{Joseph~M Hellerstein},
  \bibinfo{person}{Peter~J Haas}, {and} \bibinfo{person}{Helen~J Wang}.}
  \bibinfo{year}{1997}\natexlab{}.
\newblock \showarticletitle{Online aggregation}. In
  \bibinfo{booktitle}{\emph{Proceedings of the 1997 ACM SIGMOD international
  conference on Management of data}}. \bibinfo{pages}{171--182}.
\newblock


\bibitem[Hilprecht et~al\mbox{.}(2019)]%
        {hilprecht2019deepdb}
\bibfield{author}{\bibinfo{person}{Benjamin Hilprecht},
  \bibinfo{person}{Andreas Schmidt}, \bibinfo{person}{Moritz Kulessa},
  \bibinfo{person}{Alejandro Molina}, \bibinfo{person}{Kristian Kersting},
  {and} \bibinfo{person}{Carsten Binnig}.} \bibinfo{year}{2019}\natexlab{}.
\newblock \showarticletitle{Deepdb: Learn from data, not from queries!}
\newblock \bibinfo{journal}{\emph{arXiv preprint arXiv:1909.00607}}
  (\bibinfo{year}{2019}).
\newblock


\bibitem[Jermaine et~al\mbox{.}(2008)]%
        {jermaine2008scalable}
\bibfield{author}{\bibinfo{person}{Chris Jermaine},
  \bibinfo{person}{Subramanian Arumugam}, \bibinfo{person}{Abhijit Pol}, {and}
  \bibinfo{person}{Alin Dobra}.} \bibinfo{year}{2008}\natexlab{}.
\newblock \showarticletitle{Scalable approximate query processing with the dbo
  engine}.
\newblock \bibinfo{journal}{\emph{ACM Transactions on Database Systems (TODS)}}
  \bibinfo{volume}{33}, \bibinfo{number}{4} (\bibinfo{year}{2008}),
  \bibinfo{pages}{1--54}.
\newblock


\bibitem[Jermaine et~al\mbox{.}(2005)]%
        {jermaine2005disk}
\bibfield{author}{\bibinfo{person}{Christopher Jermaine}, \bibinfo{person}{Alin
  Dobra}, \bibinfo{person}{Subramanian Arumugam}, \bibinfo{person}{Shantanu
  Joshi}, {and} \bibinfo{person}{Abhijit Pol}.}
  \bibinfo{year}{2005}\natexlab{}.
\newblock \showarticletitle{A disk-based join with probabilistic guarantees}.
  In \bibinfo{booktitle}{\emph{Proceedings of the 2005 ACM SIGMOD international
  conference on Management of data}}. \bibinfo{pages}{563--574}.
\newblock


\bibitem[Jermaine et~al\mbox{.}(2006)]%
        {jermaine2006sort}
\bibfield{author}{\bibinfo{person}{Christopher Jermaine}, \bibinfo{person}{Alin
  Dobra}, \bibinfo{person}{Subramanian Arumugam}, \bibinfo{person}{Shantanu
  Joshi}, {and} \bibinfo{person}{Abhijit Pol}.}
  \bibinfo{year}{2006}\natexlab{}.
\newblock \showarticletitle{The sort-merge-shrink join}.
\newblock \bibinfo{journal}{\emph{ACM Transactions on Database Systems (TODS)}}
  \bibinfo{volume}{31}, \bibinfo{number}{4} (\bibinfo{year}{2006}),
  \bibinfo{pages}{1382--1416}.
\newblock


\bibitem[Joshi and Jermaine(2008)]%
        {joshi2008robust}
\bibfield{author}{\bibinfo{person}{Shantanu Joshi} {and}
  \bibinfo{person}{Christopher Jermaine}.} \bibinfo{year}{2008}\natexlab{}.
\newblock \showarticletitle{Robust stratified sampling plans for low
  selectivity queries}. In \bibinfo{booktitle}{\emph{2008 IEEE 24th
  International Conference on Data Engineering}}. IEEE,
  \bibinfo{pages}{199--208}.
\newblock


\bibitem[Kandula(2016)]%
        {kandula2016errata}
\bibfield{author}{\bibinfo{person}{Srikanth Kandula}.}
  \bibinfo{year}{2016}\natexlab{}.
\newblock \showarticletitle{Errata and Proofs for Quickr}.
\newblock \bibinfo{journal}{\emph{Technical report}} (\bibinfo{year}{2016}).
\newblock


\bibitem[Kandula et~al\mbox{.}(2016)]%
        {kandula2016quickr}
\bibfield{author}{\bibinfo{person}{Srikanth Kandula}, \bibinfo{person}{Anil
  Shanbhag}, \bibinfo{person}{Aleksandar Vitorovic}, \bibinfo{person}{Matthaios
  Olma}, \bibinfo{person}{Robert Grandl}, \bibinfo{person}{Surajit Chaudhuri},
  {and} \bibinfo{person}{Bolin Ding}.} \bibinfo{year}{2016}\natexlab{}.
\newblock \showarticletitle{Quickr: Lazily approximating complex adhoc queries
  in bigdata clusters}. In \bibinfo{booktitle}{\emph{Proceedings of the 2016
  international conference on management of data}}. \bibinfo{pages}{631--646}.
\newblock


\bibitem[Kraska(2017)]%
        {kraska2017approximate}
\bibfield{author}{\bibinfo{person}{Tim Kraska}.}
  \bibinfo{year}{2017}\natexlab{}.
\newblock \showarticletitle{Approximate query processing for interactive data
  science}. In \bibinfo{booktitle}{\emph{Proceedings of the 2017 ACM
  International Conference on Management of Data}}. \bibinfo{pages}{525--525}.
\newblock


\bibitem[Kulessa et~al\mbox{.}(2018)]%
        {kulessa2018model}
\bibfield{author}{\bibinfo{person}{Moritz Kulessa}, \bibinfo{person}{Alejandro
  Molina}, \bibinfo{person}{Carsten Binnig}, \bibinfo{person}{Benjamin
  Hilprecht}, {and} \bibinfo{person}{Kristian Kersting}.}
  \bibinfo{year}{2018}\natexlab{}.
\newblock \showarticletitle{Model-based approximate query processing}.
\newblock \bibinfo{journal}{\emph{arXiv preprint arXiv:1811.06224}}
  (\bibinfo{year}{2018}).
\newblock


\bibitem[Laptev et~al\mbox{.}(2012)]%
        {laptev2012early}
\bibfield{author}{\bibinfo{person}{Nikolay Laptev}, \bibinfo{person}{Kai Zeng},
  {and} \bibinfo{person}{Carlo Zaniolo}.} \bibinfo{year}{2012}\natexlab{}.
\newblock \showarticletitle{Early accurate results for advanced analytics on
  mapreduce}.
\newblock \bibinfo{journal}{\emph{arXiv preprint arXiv:1207.0142}}
  (\bibinfo{year}{2012}).
\newblock


\bibitem[Li et~al\mbox{.}(2016a)]%
        {li2016wander0}
\bibfield{author}{\bibinfo{person}{Feifei Li}, \bibinfo{person}{Bin Wu},
  \bibinfo{person}{Ke Yi}, {and} \bibinfo{person}{Zhuoyue Zhao}.}
  \bibinfo{year}{2016}\natexlab{a}.
\newblock \showarticletitle{Wander join: Online aggregation for joins}. In
  \bibinfo{booktitle}{\emph{Proceedings of the 2016 International Conference on
  Management of Data}}. \bibinfo{pages}{2121--2124}.
\newblock


\bibitem[Li et~al\mbox{.}(2016b)]%
        {li2016wander}
\bibfield{author}{\bibinfo{person}{Feifei Li}, \bibinfo{person}{Bin Wu},
  \bibinfo{person}{Ke Yi}, {and} \bibinfo{person}{Zhuoyue Zhao}.}
  \bibinfo{year}{2016}\natexlab{b}.
\newblock \showarticletitle{Wander join: Online aggregation via random walks}.
  In \bibinfo{booktitle}{\emph{Proceedings of the 2016 International Conference
  on Management of Data}}. \bibinfo{pages}{615--629}.
\newblock


\bibitem[Li et~al\mbox{.}(2019)]%
        {li2019wander}
\bibfield{author}{\bibinfo{person}{Feifei Li}, \bibinfo{person}{Bin Wu},
  \bibinfo{person}{Ke Yi}, {and} \bibinfo{person}{Zhuoyue Zhao}.}
  \bibinfo{year}{2019}\natexlab{}.
\newblock \showarticletitle{Wander join and XDB: online aggregation via random
  walks}.
\newblock \bibinfo{journal}{\emph{ACM Transactions on Database Systems (TODS)}}
  \bibinfo{volume}{44}, \bibinfo{number}{1} (\bibinfo{year}{2019}),
  \bibinfo{pages}{1--41}.
\newblock


\bibitem[Li and Li(2018)]%
        {li2018approximate}
\bibfield{author}{\bibinfo{person}{Kaiyu Li} {and} \bibinfo{person}{Guoliang
  Li}.} \bibinfo{year}{2018}\natexlab{}.
\newblock \showarticletitle{Approximate query processing: What is new and where
  to go?}
\newblock \bibinfo{journal}{\emph{Data Science and Engineering}}
  \bibinfo{volume}{3}, \bibinfo{number}{4} (\bibinfo{year}{2018}),
  \bibinfo{pages}{379--397}.
\newblock


\bibitem[Li et~al\mbox{.}(2018)]%
        {li_2018}
\bibfield{author}{\bibinfo{person}{Yun Li}, \bibinfo{person}{Yanlong Wen},
  {and} \bibinfo{person}{Xiaojie Yuan}.} \bibinfo{year}{2018}\natexlab{}.
\newblock \showarticletitle{Online Aggregation: A Review}. In
  \bibinfo{booktitle}{\emph{Web Information Systems and Applications}},
  \bibfield{editor}{\bibinfo{person}{Xiaofeng Meng}, \bibinfo{person}{Ruixuan
  Li}, \bibinfo{person}{Kanliang Wang}, \bibinfo{person}{Baoning Niu},
  \bibinfo{person}{Xin Wang}, {and} \bibinfo{person}{Gansen Zhao}} (Eds.).
  \bibinfo{publisher}{Springer International Publishing},
  \bibinfo{address}{Cham}, \bibinfo{pages}{103--114}.
\newblock


\bibitem[Luo et~al\mbox{.}(2002)]%
        {luo2002scalable}
\bibfield{author}{\bibinfo{person}{Gang Luo}, \bibinfo{person}{Curt~J Ellmann},
  \bibinfo{person}{Peter~J Haas}, {and} \bibinfo{person}{Jeffrey~F Naughton}.}
  \bibinfo{year}{2002}\natexlab{}.
\newblock \showarticletitle{A scalable hash ripple join algorithm}. In
  \bibinfo{booktitle}{\emph{Proceedings of the 2002 ACM SIGMOD international
  conference on Management of data}}. \bibinfo{pages}{252--262}.
\newblock


\bibitem[McDiarmid(1998)]%
        {mcdiarmid1998concentration}
\bibfield{author}{\bibinfo{person}{Colin McDiarmid}.}
  \bibinfo{year}{1998}\natexlab{}.
\newblock \showarticletitle{Concentration, Probabilistic methods for
  algorithmic discrete mathematics, 195--248}.
\newblock \bibinfo{journal}{\emph{Algorithms Combin}}  \bibinfo{volume}{16}
  (\bibinfo{year}{1998}).
\newblock


\bibitem[Mozafari(2017)]%
        {mozafari2017approximate}
\bibfield{author}{\bibinfo{person}{Barzan Mozafari}.}
  \bibinfo{year}{2017}\natexlab{}.
\newblock \showarticletitle{Approximate query engines: Commercial challenges
  and research opportunities}. In \bibinfo{booktitle}{\emph{Proceedings of the
  2017 ACM International Conference on Management of Data}}.
  \bibinfo{pages}{521--524}.
\newblock


\bibitem[Olken and Rotem(1986)]%
        {olken1986simple}
\bibfield{author}{\bibinfo{person}{Frank Olken} {and} \bibinfo{person}{Doron
  Rotem}.} \bibinfo{year}{1986}\natexlab{}.
\newblock \showarticletitle{Simple random sampling from relational databases}.
\newblock  (\bibinfo{year}{1986}).
\newblock


\bibitem[Olken and Rotem(1995)]%
        {olken1995random}
\bibfield{author}{\bibinfo{person}{Frank Olken} {and} \bibinfo{person}{Doron
  Rotem}.} \bibinfo{year}{1995}\natexlab{}.
\newblock \showarticletitle{Random sampling from databases: a survey}.
\newblock \bibinfo{journal}{\emph{Statistics and Computing}}
  \bibinfo{volume}{5}, \bibinfo{number}{1} (\bibinfo{year}{1995}),
  \bibinfo{pages}{25--42}.
\newblock


\bibitem[Pansare et~al\mbox{.}(2011)]%
        {pansare2011online}
\bibfield{author}{\bibinfo{person}{Niketan Pansare}, \bibinfo{person}{Vinayak
  Borkar}, \bibinfo{person}{Chris Jermaine}, {and} \bibinfo{person}{Tyson
  Condie}.} \bibinfo{year}{2011}\natexlab{}.
\newblock \showarticletitle{Online aggregation for large mapreduce jobs}.
\newblock \bibinfo{journal}{\emph{Proceedings of the VLDB Endowment}}
  \bibinfo{volume}{4}, \bibinfo{number}{11} (\bibinfo{year}{2011}),
  \bibinfo{pages}{1135--1145}.
\newblock


\bibitem[Park et~al\mbox{.}(2018)]%
        {park2018verdictdb}
\bibfield{author}{\bibinfo{person}{Yongjoo Park}, \bibinfo{person}{Barzan
  Mozafari}, \bibinfo{person}{Joseph Sorenson}, {and} \bibinfo{person}{Junhao
  Wang}.} \bibinfo{year}{2018}\natexlab{}.
\newblock \showarticletitle{Verdictdb: Universalizing approximate query
  processing}. In \bibinfo{booktitle}{\emph{Proceedings of the 2018
  International Conference on Management of Data}}.
  \bibinfo{pages}{1461--1476}.
\newblock


\bibitem[Park et~al\mbox{.}(2017)]%
        {park2017database}
\bibfield{author}{\bibinfo{person}{Yongjoo Park}, \bibinfo{person}{Ahmad~Shahab
  Tajik}, \bibinfo{person}{Michael Cafarella}, {and} \bibinfo{person}{Barzan
  Mozafari}.} \bibinfo{year}{2017}\natexlab{}.
\newblock \showarticletitle{Database learning: Toward a database that becomes
  smarter every time}. In \bibinfo{booktitle}{\emph{Proceedings of the 2017 ACM
  International Conference on Management of Data}}. \bibinfo{pages}{587--602}.
\newblock


\bibitem[Piatetsky-Shapiro and Connell(1984)]%
        {piatetsky1984accurate}
\bibfield{author}{\bibinfo{person}{Gregory Piatetsky-Shapiro} {and}
  \bibinfo{person}{Charles Connell}.} \bibinfo{year}{1984}\natexlab{}.
\newblock \showarticletitle{Accurate estimation of the number of tuples
  satisfying a condition}.
\newblock \bibinfo{journal}{\emph{ACM Sigmod Record}} \bibinfo{volume}{14},
  \bibinfo{number}{2} (\bibinfo{year}{1984}), \bibinfo{pages}{256--276}.
\newblock


\bibitem[Qin and Rusu(2013)]%
        {qin2013parallel}
\bibfield{author}{\bibinfo{person}{Chengjie Qin} {and} \bibinfo{person}{Florin
  Rusu}.} \bibinfo{year}{2013}\natexlab{}.
\newblock \showarticletitle{Parallel online aggregation in action}. In
  \bibinfo{booktitle}{\emph{Proceedings of the 25th International Conference on
  Scientific and Statistical Database Management}}. \bibinfo{pages}{1--4}.
\newblock


\bibitem[Qin and Rusu(2014)]%
        {qin2014pf}
\bibfield{author}{\bibinfo{person}{Chengjie Qin} {and} \bibinfo{person}{Florin
  Rusu}.} \bibinfo{year}{2014}\natexlab{}.
\newblock \showarticletitle{PF-OLA: a high-performance framework for parallel
  online aggregation}.
\newblock \bibinfo{journal}{\emph{Distributed and Parallel Databases}}
  \bibinfo{volume}{32}, \bibinfo{number}{3} (\bibinfo{year}{2014}),
  \bibinfo{pages}{337--375}.
\newblock


\bibitem[Singh and Singh(2018)]%
        {singh2018sampling}
\bibfield{author}{\bibinfo{person}{Sarjinder Singh} {and} \bibinfo{person}{S
  Singh}.} \bibinfo{year}{2018}\natexlab{}.
\newblock \bibinfo{title}{Sampling techniques}.
\newblock
\newblock


\bibitem[Song et~al\mbox{.}(2018)]%
        {song2018approximate}
\bibfield{author}{\bibinfo{person}{Guangxuan Song}, \bibinfo{person}{Wenwen
  Qu}, \bibinfo{person}{Xiaojie Liu}, {and} \bibinfo{person}{Xiaoling Wang}.}
  \bibinfo{year}{2018}\natexlab{}.
\newblock \showarticletitle{Approximate calculation of window aggregate
  functions via global random sample}.
\newblock \bibinfo{journal}{\emph{Data Science and Engineering}}
  \bibinfo{volume}{3}, \bibinfo{number}{1} (\bibinfo{year}{2018}),
  \bibinfo{pages}{40--51}.
\newblock


\bibitem[Thirumuruganathan et~al\mbox{.}(2020)]%
        {thirumuruganathan2020approximate}
\bibfield{author}{\bibinfo{person}{Saravanan Thirumuruganathan},
  \bibinfo{person}{Shohedul Hasan}, \bibinfo{person}{Nick Koudas}, {and}
  \bibinfo{person}{Gautam Das}.} \bibinfo{year}{2020}\natexlab{}.
\newblock \showarticletitle{Approximate query processing for data exploration
  using deep generative models}. In \bibinfo{booktitle}{\emph{2020 IEEE 36th
  International Conference on Data Engineering (ICDE)}}. IEEE,
  \bibinfo{pages}{1309--1320}.
\newblock


\bibitem[Vengerov et~al\mbox{.}(2015)]%
        {vengerov2015join}
\bibfield{author}{\bibinfo{person}{David Vengerov},
  \bibinfo{person}{Andre~Cavalheiro Menck}, \bibinfo{person}{Mohamed Zait},
  {and} \bibinfo{person}{Sunil~P Chakkappen}.} \bibinfo{year}{2015}\natexlab{}.
\newblock \showarticletitle{Join size estimation subject to filter conditions}.
\newblock \bibinfo{journal}{\emph{Proceedings of the VLDB Endowment}}
  \bibinfo{volume}{8}, \bibinfo{number}{12} (\bibinfo{year}{2015}),
  \bibinfo{pages}{1530--1541}.
\newblock


\bibitem[Wang et~al\mbox{.}(2014)]%
        {wang2014oats}
\bibfield{author}{\bibinfo{person}{Yuxiang Wang}, \bibinfo{person}{Junzhou
  Luo}, \bibinfo{person}{Aibo Song}, {and} \bibinfo{person}{Fang Dong}.}
  \bibinfo{year}{2014}\natexlab{}.
\newblock \showarticletitle{OATS: online aggregation with two-level sharing
  strategy in cloud}.
\newblock \bibinfo{journal}{\emph{Distributed and Parallel Databases}}
  \bibinfo{volume}{32}, \bibinfo{number}{4} (\bibinfo{year}{2014}),
  \bibinfo{pages}{467--505}.
\newblock


\bibitem[Wu et~al\mbox{.}(2009)]%
        {wu2009distributed}
\bibfield{author}{\bibinfo{person}{Sai Wu}, \bibinfo{person}{Shouxu Jiang},
  \bibinfo{person}{Beng~Chin Ooi}, {and} \bibinfo{person}{Kian-Lee Tan}.}
  \bibinfo{year}{2009}\natexlab{}.
\newblock \showarticletitle{Distributed online aggregations}.
\newblock \bibinfo{journal}{\emph{Proceedings of the VLDB Endowment}}
  \bibinfo{volume}{2}, \bibinfo{number}{1} (\bibinfo{year}{2009}),
  \bibinfo{pages}{443--454}.
\newblock


\bibitem[Wu et~al\mbox{.}(2010)]%
        {wu2010continuous}
\bibfield{author}{\bibinfo{person}{Sai Wu}, \bibinfo{person}{Beng~Chin Ooi},
  {and} \bibinfo{person}{Kian-Lee Tan}.} \bibinfo{year}{2010}\natexlab{}.
\newblock \showarticletitle{Continuous sampling for online aggregation over
  multiple queries}. In \bibinfo{booktitle}{\emph{Proceedings of the 2010 ACM
  SIGMOD International Conference on Management of data}}.
  \bibinfo{pages}{651--662}.
\newblock


\bibitem[Zeng et~al\mbox{.}(2015)]%
        {zeng2015g}
\bibfield{author}{\bibinfo{person}{Kai Zeng}, \bibinfo{person}{Sameer Agarwal},
  \bibinfo{person}{Ankur Dave}, \bibinfo{person}{Michael Armbrust}, {and}
  \bibinfo{person}{Ion Stoica}.} \bibinfo{year}{2015}\natexlab{}.
\newblock \showarticletitle{G-ola: Generalized on-line aggregation for
  interactive analysis on big data}. In \bibinfo{booktitle}{\emph{Proceedings
  of the 2015 ACM SIGMOD International Conference on Management of Data}}.
  \bibinfo{pages}{913--918}.
\newblock


\bibitem[Zeng et~al\mbox{.}(2014)]%
        {zeng2014analytical}
\bibfield{author}{\bibinfo{person}{Kai Zeng}, \bibinfo{person}{Shi Gao},
  \bibinfo{person}{Barzan Mozafari}, {and} \bibinfo{person}{Carlo Zaniolo}.}
  \bibinfo{year}{2014}\natexlab{}.
\newblock \showarticletitle{The analytical bootstrap: a new method for fast
  error estimation in approximate query processing}. In
  \bibinfo{booktitle}{\emph{Proceedings of the 2014 ACM SIGMOD international
  conference on Management of data}}. \bibinfo{pages}{277--288}.
\newblock


\bibitem[Zhang et~al\mbox{.}(2016)]%
        {zhang2016sapprox}
\bibfield{author}{\bibinfo{person}{Xuhong Zhang}, \bibinfo{person}{Jun Wang},
  {and} \bibinfo{person}{Jiangling Yin}.} \bibinfo{year}{2016}\natexlab{}.
\newblock \showarticletitle{Sapprox: Enabling efficient and accurate
  approximations on sub-datasets with distribution-aware online sampling}.
\newblock \bibinfo{journal}{\emph{Proceedings of the VLDB Endowment}}
  \bibinfo{volume}{10}, \bibinfo{number}{3} (\bibinfo{year}{2016}),
  \bibinfo{pages}{109--120}.
\newblock


\end{thebibliography}

\end{document}